\documentclass[10pt,conference]{IEEEtran}

\makeatletter
\def\ps@headings{%
\def\@oddhead{\mbox{}\scriptsize\rightmark \hfil \thepage}%
\def\@evenhead{\scriptsize\thepage \hfil \leftmark\mbox{}}%
\def\@oddfoot{}%
\def\@evenfoot{}}
\makeatother
\usepackage{cite}
\usepackage{amsmath,amssymb,amsfonts}
\usepackage{algorithmic}
\usepackage{graphicx}
\usepackage{textcomp}
\usepackage{xcolor}
\usepackage{hyperref}
\usepackage{csvsimple}
\def\BibTeX{{\rm B\kern-.05em{\sc i\kern-.025em b}\kern-.08em
    T\kern-.1667em\lower.7ex\hbox{E}\kern-.125emX}}

\usepackage[textsize=tiny,disable]{todonotes} % Todonotes is off

\newcommand{\dsdn}{\textsc{DocSDN}}

\newcommand{\comment}[1]{}

\begin{document}
\title{DOCSDN: Dynamic and Optimal Configuration of Software-Defined Networks}

\author{\IEEEauthorblockN{ Timothy Curry, Devon Callahan, Benjamin Fuller and Laurent Michel}
\IEEEauthorblockA{\textit{Department of Computer Science and Engineering} \\
\textit{University of Connecticut}
Storrs, CT USA \\
\{timothy.curry,devon.callahan,benjamin.fuller,laurent.michel\}@uconn.edu}
}
\maketitle
%!TEX root = main.tex
\begin{abstract}
Networks are designed with  functionality, security, performance, and
cost in mind. Tools exist to check or optimize individual properties
of a network.  These properties may conflict, so it is not always
possible to run these tools in series to find a configuration that
meets all requirements.  This leads to network administrators manually
searching for a configuration.   

This need not be the case. In this paper, we introduce a layered
framework for optimizing network configuration for functional and
security requirements.  Our framework is able to output configurations
that meet reachability,  bandwidth, and risk requirements.  Each layer
of our framework optimizes over a single property.  A lower layer can
constrain the search problem of a higher layer allowing the framework
to converge on a joint solution. 

Our approach has the most promise for software-defined networks which
can easily reconfigure their logical configuration.  Our approach is
validated with experiments over the fat tree topology, which is 
commonly used in data center networks.  Search terminates in between 1-5 minutes in experiments.
Thus, our solution can propose new configurations for short term
events such as defending against a focused network attack. 

\end{abstract}

%!TEX root = main.tex

\section{Introduction}

Network configuration is a crucial task in any enterprise.  Administrators balance functionality, performance, security, cost and other
industry specific requirements.  The resulting configuration is
subject to periodic analysis and redesign due to red team
recommendations, emerging threats, and changing
priorities.  
Tools assist administrators with this complex task: existing work
assesses network reachability~\cite{khurshid2012veriflow}, wireless
conflicts~\cite{neves2016selfnet}, network security
risk~\cite{Schneier2,yu2018deploying}, and load balancing
\cite{wang2011openflow,skowyra2014verification}.  These tools assess
the quality of a potential configuration.  Unfortunately, current
tools suffer from three limitations:

\begin{enumerate}
\itemsep0em
\item Most tools assess whether a single property is satisfied, making
  no recommendation if the property is not satisfied.  This leaves IT
  personnel with the task of deciding how to change the network. 
\item Networks are assessed with respect to an individual goal at a
  time.  This means a change to satisfy a single property may break
  another property.  There is no guidance for personnel on how to
  design a network that meets the complex and often conflicting
  network requirements. 
\item These tools do not react to changing external information such
  as the publication of a new security vulnerability. 
\end{enumerate}

\paragraph{Our Contribution}
This work introduces a new optimization framework that finds network
configurations that satisfy multiple (conflicting) requirements.   We
focus on data center networks (DCN) that use software defined
networking (SDN).  Background on these settings is in
Section~\ref{sec:background}.  Our framework is called \dsdn{} (Dynamic
and Optimal Configuration of Software-Defined Networks).  

\dsdn{} searches for network configurations that simultaneously satisfy
multiple properties.  \dsdn{} is organized into layers that consider
different properties.  The core of \dsdn{} is a multistage optimization
that decouples search on ``orthogonal'' concerns.  The majority of the
technical work is to effectively separate 
concerns so the optimization problems remain tractable.   
Our framework is designed to continually produce network
configurations based on changing requirements and threats.  It frees
IT personnel from the complex question of how to satisfy multiple
requirements and can quickly incorporate new threat information.   

\dsdn{} focuses on achieving functional requirements (such as
network reachability and flow satisfaction) and limiting security risk
(such as isolating high risk nodes and nodes under denial of service
attack).  Naturally, other layers such as performance or cost can be
  incorporated.  The search for a good
configuration could be organized in many ways.  State-of-the-art approaches assess
different properties in isolation, frustrating search for a
solution that satisfies all requirements.  Ideally, a
framework should search for a configuration that simultaneously
satisfies all requirements.  This extreme is unlikely to be tractable
on all but the smallest networks.  \dsdn{} mediates between these
approaches separating the functional and security search problems but
introducing a feedback loop between the two search problems based on \emph{cuts}.

In the proposed organization the functional layer is ``above'' the
security layer.  Through the feedback loop, the security layer
describes a problematic part of the network to the functional layer.
The functional layer then refines its model and searches for a
functional configuration that satisfies an additional \emph{constraint}.  This has the 
effect of blocking the problematic part of the configuration.  Currently,
the feedback signal is a pair of nodes that should not be
proximate in the network. After
multiple iterations the two layers jointly produce a solution that
optimizes the SDN configuration both with respect to functionality and
security risks.  

\dsdn{}
provides solutions of improving quality before the final solution.  
Thus, the network can be reconfigured once the objective
improves on the current configuration by a large enough amount (to
justify the cost/impact of reconfiguration).

While the underlying optimization problems are NP-hard, optimization
technology has seen tremendous advances in performance during the past
few decades.  Since 1991, mathematical programming solvers
have delivered speedups of 11 orders of
magnitude~\cite{Bixby2000,ASS2015}.
The advent of hybrid techniques such as Benders
decomposition~\cite{Benders1962,Hooker95,HookerBook,Codato2006} and
column generation~\cite{barnhart1998branch,Hijazi:2015,Lam:2016} (aka,
Dantzig-Wolfe decomposition~\cite{Dantzig:1960}) made it possible to
harness huge problems thanks to on-demand generation of macroscopic
variables and the dynamic addition of critical constraints. Large
Neighborhood Search~\cite{Shaw98} further contributed to delivering
high-quality solution within constrained time budgets. 

These techniques 
are beginning to see adoption in network security.
Yu et al. recently applied stochastic optimization with
Bender's decomposition to assess network risk under uncertainty for
IoT devices~\cite{yu2018deploying}.
They used Bender's decomposition on a scenario-based stochastic
optimization model to produce a parent problem that chooses a
deployment plan while children are
concerned with \emph{choosing} the optimal nodes to serve the
demands in individual scenarios. In comparison, our approach addresses
\emph{both functional and security requirements}. It relies on
Bender's cuts from the security layer (child) to rule out vulnerable 
functional plans whose routing paths fail to adequately minimize risks
and maximize served clients. 
We now briefly describe the framework (a
formal description is in Section~\ref{sec:methods}) and present an
illustrative example. 

\paragraph{Overview of \dsdn}

\begin{figure}
  \begin{center}
%    \vspace{.21in}
    \includegraphics[width=0.45\textwidth]{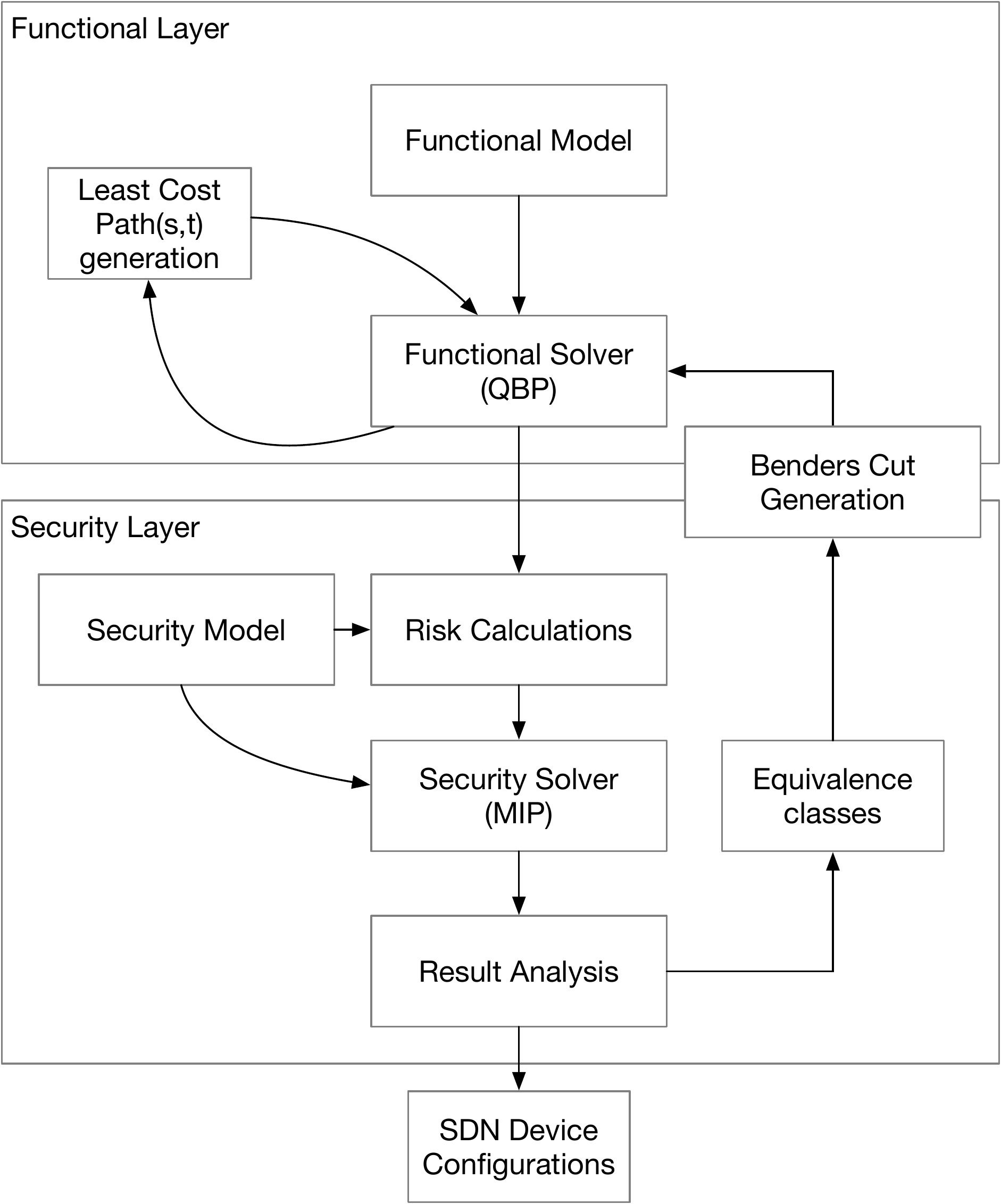}
%    \vspace{-4mm}
\end{center}
\caption{\dsdn{} Framework. {\small A layered decomposition that
    breaks down configuration synthesis into functional and security
    layers.}}
\label{fig:hierarchy}
\end{figure}

Figure~\ref{fig:hierarchy} presents an overview of the framework.
The functional layer takes as input a \emph{Functional Model} that
describes the network including the physical topology, link speeds,
the allowable communication patterns and the demand requirements.
Network reachability begins with a priming procedure that generates
the $k$-least cost paths to the optimizer for each source/destination
pair in the demand requirements.  The objective for 
the functional layer is to find a logical topology (a collection of
routed paths) that meets all demand requirements while favoring 
shorter length routing paths and load balancing. 
The program is formulated as
quadratic binary program (\emph{QBP}).  The solution as determined by
the functional layer is passed to the security layer. 

The output of the functional layer and a \emph{security model} are the
input for the security layer. The current configuration is fed to a
module that uses risk assessments for the individual network
devices (obtained for example using a vulnerability database) to assess the overall risk of the
entire configuration.
In our current implementation this risk
calculation is based on a simple risk propagation model where a path's risk
is based on the risk of nodes on the path and close to the source and sink.  
The security
layer can deploy firewalls and deep packet inspection as network defenses. Since these mechanisms affect route capacity, the
security layer has a dual objective function: 1) maximizing the
functional objective and 2) minimizing security risk.  The security
objective is formulated as a mixed integer program (\emph{MIP}).  When
the security search completes, it proposes nodes to the functional
layer that should be separated.  As an example, a high value node with
low risk may be placed in a different (virtual) LAN than a high risk node.
These \emph{Benders cuts} are designed to entice a better logical topology
from a subsequent iteration in the functional layer.
This feedback loop between the two layers can iterate multiple
times.  When no further cuts are available, the overall output is a
set of configuration rules.

%!TEX root = main.tex

\paragraph{An example configuration}

\begin{figure}[t]
\centering
%\vspace{.06in}
\includegraphics[scale=.7]{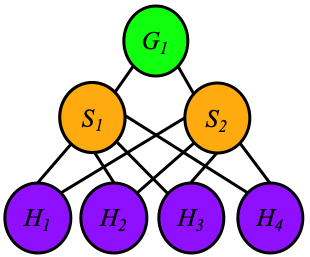}
\caption{Toy network example with a single gateway device $G_1$, two
  intermediate switches $S_1$ and $S_2$ and four hosts.  We assume the
  switches are physically connected to all hosts.} 
\label{fig:basic attack}
\end{figure}

This section describes an application of our framework to
automatically respond to a distributed denial of service (DDoS)
attack.  Current DDoS attacks demonstrate peak volume of 1
Tbps~\cite{skottler2018}. Many DDoS defense techniques require changes
to the network behavior by rate limiting, filtering, or reconfiguring
the network (see
\cite{Ioannidis01pushback:router-based,ioannidis2002implementing,mirkovic2004taxonomy,peng2007survey,zargar2013survey}).
Recent techniques~\cite{fayaz2015bohatei} leverage SDNs to react to
DDoS attacks in a dynamic and flexible manner.
We show how such a response would work in our framework using a toy
network illustrated in Figure~\ref{fig:basic attack}. A more realistic network and the
framework's response are described in Section~\ref{sec:eval}.  We stress that
DDoS attacks are often short in timescale making human diagnosis and
reaction costly or impractical.

Consider a focused DDoS attack against a number of services in an
enterprise but not the entirety of its publicly accessible address
space. (The Great Cannon's attack against GreatFire targeted two
specific Github repositories~\cite{marczak2015china}.) We assume a
service hosted by $H_1$ is targeted, while services on $H_2, H_3$ and
$H_4$ are not.
 
Recall, the functional layer establishes a logical topology
(forwarding rules) while the security layer adds network defenses
(packet inspection modules and firewall rules).  We elide how the
attack is detected and assume it increases the risk score for $H_1$ in
the security model.

\noindent\textbf{The first iteration}
The functional layer proposes a candidate configuration where $G_1$
routes all traffic intended for $H_1$ and $H_2$ to $S_1$ which then
forwards the traffic and $G_1$ routes traffic intended for $H_3$ and
$H_4$ to $S_2$ which then forwards the traffic.  This is the first
candidate solution presented to the security layer.

Since $H_1$ is high risk the security layer proposes a firewall at $S_1$
to block all port $80$ traffic. This reduces risk at the cost of blocking all traffic to
$H_2$.  Of course, in real firewalls more fine-grained rules are
possible, this simplified example is meant to illustrate a case where
collateral damage to the functional objective is necessary to achieve
the security objective.  Since traffic is being blocked to a node with
low risk, the security layer asks the functional layer to separate
$H_1$ and $H_2$ so $H_2$ does not suffer.

\noindent\textbf{Repeated iterations}
The functional layer now has a constraint that $H_1$ and $H_2$ should
not be collocated in the network.  As such, it proposes a new
configuration with $H_1$ and $H_3$ under $S_1$ and $H_2$ and $H_4$ under
$S_2$.  This is then sent to the security layer.  The security layer
makes a similar assessment and proposes a firewall rule at $S_2$, finds 
this recommendation hurts functionality and 
requests separation of $H_1$ and $H_3$.

This process repeats with the functional layer proposing to collocate
$H_1$ and $H_4$.  The security layer similarly asks to
separate $H_1$ and $H_4$.  Finally, $H_1$ is segregated from all other
nodes.  This produces a configuration where $H_1$ is the only child of
$S_1$.  Note that having $H_2, H_3$ and $H_4$ under a single switch may
hurt performance but the effect is less than blocking traffic to one
of the nodes entirely.
\dsdn{} can then output the candidate solution as high level SDN
fragments (using a high-level language like Pyretic~\cite{reich2013modular}).
      
\noindent\textbf{Recovery}
Importantly, when the DDoS abates, $\dsdn$ automatically reruns with a
changed risk for $H_1$, outputting a binary tree.

\noindent\textbf{Organization}
The rest of the work is organized as follows:
Section~\ref{sec:background} provides background on our application
and discusses related work, Section~\ref{sec:methods} describes our
framework and accompanying optimization models,
Section~\ref{sec:setup} describes our experimental setup,
Section~\ref{sec:eval} evaluates the framework and finally
Section~\ref{sec:conclusion} concludes.

%%% Local Variables:
%%% mode: latex
%%% TeX-master: "main"
%%% End:

%%% Local Variables:
%%% mode: latex
%%% TeX-master: "main"
%%% End:

%!TEX root = main.tex

\section{Background and Related Work}
\label{sec:background}
\label{ssec:setting}
Data Center Networks (DCN) host, process and analyze data in financial, entertainment, medical, government and military sectors. The services provided by DCNs must be reliable, accurate and timely.  Services provided by DCNs (and the corresponding traffic) are heterogeneous.   
The network must adapt to changing priorities and requirements while protecting from emerging threats.
They scale to thousands of
servers linked through various interconnects.   
Protocols used for these services are split roughly 60 percent
web~(HTTP/HTPS) and 40 percent file storage
(SMB/AFS)~\cite{Benson:2010:NTC:1879141.1879175}.
The interdependence of device configurations make modifying any single configuration difficult and possibly dangerous for network health.  A seemingly simple update can cause significant collateral damage and unintended consequences.

Simultaneously, the network fabric is changing with the advent of Software Defined Networking (SDN)~\cite{kreutz2015software}. SDNs are flexible and programmable networks that can adapt to emergent functional or performance requirements.   Openflow~\cite{mckeown2008openflow} is a common open source software stack.  Researchers have proposed high-level languages and compilers~\cite{foster2011frenetic,kim2015kinetic,reich2013modular,beckett2017network} that bridge the semantic gap between network administrators and the configuration languages used by SDN devices. These languages focus on
  \emph{compositional} and \emph{parametric} SDN software modules that
  execute specific micro-functions (e.g., packet forwarding, dropping,
  routing, etc.).  The use of a high level language is prompted by a
  desire to be able to \emph{select}, \emph{instantiate} and
  \emph{compose} SDN modules with guarantees.

%\paragraph{Prior Work}
\label{sec:prior work}
Our framework is intended to be modular and allow integration of prior work on evaluating network configurations. As such there is a breadth of relevant work.  Due to space constraints we focus on the most relevant works.  In the conclusion we elaborate on the characteristics needed to integrate a prior assessment tool into our framework (see Section~\ref{sec:conclusion}). 

\textbf{Measuring Network Risk}
Known threats against computer systems are maintained by governments and industry.  Common Vulnerabilities and Exposures (CVE) is a publicly available dictionary including an identifier and description of known vulnerabilities~\cite{website:CVE}, CVE does not provide a severity score or priority ranking for vulnerabilities.  The US National Vulnerability Database (NVD)~\cite{website:NVD} is  provided by the US National Institute of Standards and Technology (NIST).  The NVD augments the CVE, adding severity scores and impact ratings for vulnerabilities in the CVE.

There are many mechanisms for measuring the security risk on a network~\cite{stoneburner2002sp,jansen2010directions,stolfo2011measuring,lippmann2012continuous,cherdantseva2016review}.   Lippmann et al. present a network security model which computes risk based on a list of the most current threats~\cite{lippmann2016threat}.  This model implements a cycle of observe network state, compute risk, prioritize risk, and mitigate the risk.  

  This loop is often codified into an \emph{attack graph}~\cite{Schneier2,ingols2006practical,kaynar2016taxonomy}.  Attack graphs try to model the most likely paths that an attacker could use to penetrate a network.  Attack Graphs often leverage one or more of the aforementioned vulnerability assessment tools as input, combined with a network topology and device software configurations to generate the graph. Current attack graph technologies provide recommendations to network administrators that effectively remove edges from the graph and trigger a re-evaluation of the utility for the attacker. To the best of our knowledge, current practice does not leverage network risk measurement into constraints used for the generation of new configurations. 

\textbf{Network Reachability}
The expansion of SDN has aided the applicability of formal verification to computer networks.  Prior to SDN, the lack of clear separation between the data and control plain created an intractable problem when considering a network of any scale.  Bounded model checking using SAT and SMT solvers~\cite{zhang2013sat,beckett2017general} can currently verify reachability properties in networks with several thousands of nodes.

\textbf{Configuration Search}
 Constraint Programming (CP) was introduced in the late 1980s~\cite{Rossi:2006:HCP:1207782} and is used for scheduling~\cite{Baptiste2001}, routing, and configuration problems. Large-scale optimization problems are often decomposed including Benders~\cite{Codato2006} and Dantzig-Wolfe~\cite{Dantzig:1960}. \emph{Soft constraints} or Lagrangian relaxation  are used for over-constrained problems or when the problem is too computationally expensive. Stochastic optimization techniques have been used for many applications in resilience \cite{DBLP:conf/pscc/NagarajanYBHB16,Byeon18} and the underlying methodologies are a key part of this research. Prior work in configuration management with constraint programming~\cite{DBLP:conf/cds/CoattaN92, DBLP:journals/corr/LayeghyPP16} focused on connectivity or security.  We are not aware of any work that balances these two objectives in a meaningful way.

%%% Local Variables:
%%% mode: latex
%%% TeX-master: "main"
%%% End:

%!TEX root = main.tex
%\vspace{-.05in}
\section{Implementation}
\label{sec:methods}
Figure~\ref{fig:hierarchy} outlines the overall structure of the
\dsdn{} framework. 
layer inter-connections as well as their internals. 
The
functional layer uses a mathematical optimization model that is fed
to a quadratic mixed Boolean programming (QBP) solver alongside an initial set of
least-cost paths to be considered to service the required flows.
The security layer receives the topology chosen by the functional
layer and a security model to solve, with a mixed-integer
programming (MIP) solver, the risk minimization problem. The output can result in
low-risk flows being blocked as a consequence of deploying firewalls 
to mitigate high-risk flows.
A result analysis module then produces a set of \emph{equivalence classes} that is
sent back to the functional layer to request the separation of specific
flows that should not share paths, with the goal of minimizing 
the collateral damage to low-risk flows. These
equivalence classes generate additional constraints, known as 
Bender's cuts, that are added to the functional
solver for a new iteration. 
The remainder of this section describes the major modules in Figure~\ref{fig:hierarchy}. 

\subsection{Functional Layer}
\label{sec:fun}

%% --------------------------------------------------------------------------------
%% --------------------------------------------------------------------------------
%% --------------------------------------------------------------------------------

The mathematical optimization model in the functional layer is a
quadratic mixed binary programming model. In constraint programming the four main components are Inputs, Variables, Constraints, and an Objective function.  Inputs are  below.

\noindent\textbf{Inputs}\\
% networkDevices
${\cal N}$ -- the set of all network devices\\
${\cal E}$ -- the set of edges (pair of vertices) connecting network devices\\
${\cal T}$ -- the set of types of traffic to be routed \\
\noindent
% desiredFlows
${\cal F}$ -- the set of $(s, t, T) \in {\cal N} \times {\cal N} \times {\cal T}$
tuples defining desired traffic flows of type $T$ from source node $s$ to sink node $t$.\\ 
\noindent
% Demand
$D(f) : {\cal F} \rightarrow \mathbb{R}$ -- the actual demand for each flow $f \in {\cal F}$\\
% Class
${\cal C} \subseteq 2^{\cal N}$ -- a subset of sets of network devices\\
% EquivRules
${\cal R} \subseteq {\cal C}\times {\cal C}$ -- pairs $(c_1,c_2)$ of
equivalence classes that segregate traffic from  $c_1$ to $c_2$. \\
\noindent
% Path
${\cal P}$ -- the set of all paths\\
$P(e) : {\cal E} \rightarrow {\cal P}$ -- the set of all paths containing edge $e$\\
$P(n) : {\cal N} \rightarrow {\cal P}$ -- the set of all paths containing node $n$\\
$P(c) : {\cal C} \rightarrow {\cal P}$ -- the set of all paths containing a node in $c$\\ 
$N(p) : {\cal P} \rightarrow {\cal C}$ -- the set of nodes appearing
in path $p$\\
$P(s,t) : {\cal N\times N} \rightarrow {\cal P}$ -- the set of all paths  $s \rightarrow t$\\
$cap(e): {\cal E} \rightarrow \mathbb{R}$ -- gives the capacity of an
edge $e$.\\

\vspace{-.1in}
\noindent\textbf{Variables}\\
% isFlow
$active_{p, T} \in \{0,1\}$, -- for every path $p \in {\cal P}$
and traffic type $T\in{\cal T}$, indicates whether path $p$ carries traffic of type $T$\\
% flow
$flow_{p, T} \in \mathbb{R}_{\geq 0}$ -- for every path $p \in {\cal P}$
and traffic type $T\in{\cal T}$, amount of flow of type $T$ that is sent along path $p$\\
% equiv
$equiv_{c, n} \in \{0,1\}$ -- does node $n \in{\cal N}$ appear in an
active path together with a node in equivalence class $c$\\ 
% overlap
$share_{c_1, c_2, n}\in \{0,1\}$ --
indicates whether node $n\in {\cal N}$ appears on any active path
with nodes in classes $c_1,c_2 \in {\cal C}$. Namely,
\[
  share_{c_1,c_2,n} \Leftrightarrow n \in
  \left(\begin{array}{c}
    \left(
    \cup_{p \in P(c_1) : active_{p,*}} N(p)
    \right)\\
    \cap \\    
    \left(
    \cup_{p \in P(c_2) : active_{p,*}} N(p)
    \right)
  \end{array}\right)
\]
$active_{p,*}=1$ if there is a type $T\in{\cal T}$ where
$active_{p,T}=1$\\
% load
$load_{n} \in \mathbb{R}$ -- the amount of flow that goes through node $n$\\
% loadSquaresSum
$loadObj$ -- the sum of squares of all $load_n$ variables.\\

\vspace{-.10in}
\noindent\textbf{Constraints}\\
\begin{equation}
  \label{eq1} % edge capacity
  \sum_{p \in P(e),T \in {\cal T}}{flow_{p, T}} \leq cap(e), \:\:\forall  e \in {\cal E}
\end{equation}
\vspace{-.04in}
\begin{equation}
  \label{eq2} % demand
\sum_{p \in P(s,t)}{flow_{p,T}} \geq D(s,t,T),  \:\:\forall (s,t,T) \in {\cal F}
\end{equation}
\vspace{-.04in}
\begin{equation}
  \label{eq3} % isFlow => flow
\hspace{-.5em}active_{p,T}\!=\!1\rightarrow flow_{p,T}\geq\! 1,\forall (s,t,T) \in{\cal F},p\in P(s,t)
\end{equation}
\vspace{-.04in}
\begin{equation}
  \label{eq4} % flow
\sum_{p \in P(s,t)}{active_{p, T}} = 1, \:\: \forall (s, t, T) \in {\cal  F}
\end{equation}
\begin{equation}
  \label{eq5} % equiv
equiv_{c,n} \!=\!\!\!\!\! \bigvee_{p \in P(n) \cap P(c)} \!(active_{p, T}),\forall T\in{\cal T},n \in {\cal N},c \in {\cal C}
\end{equation}
\begin{equation}
  \label{eq6} % overlap
  \hspace{-.5em}
\!share_{c_1,c_2, n}\!=\!equiv_{c_1,n} \land equiv_{c_2,n},\forall n \in {\cal N},(c_1,c_2)\!\in\!{\cal R}
\end{equation}
%
%\vspace{.05in}
\begin{equation}
  \label{eq7} % load
load_n = \sum_{p \in P(n),T \in {\cal T}}{flow_{p, T}},\forall n \in {\cal N}
\end{equation}

Equation~\ref{eq1} enforces the edge capacity constraint to service
the demand of all paths flowing through it. Equation~\ref{eq2} ensures
that enough capacity is available 
 to meet the demand of an  $(s,t,T)$ flow. Equation~\ref{eq3} ensures that
some non-zero capacity is used if a specific path is activated
(conversely, an inactive path can only have a 0
flow). Equation~\ref{eq4} states that a single path should be chosen
to service a given flow $f \in {\cal F}$.
Equations~\ref{eq5} define the auxiliary variables $equiv_{c,n}$ as
true if and only if node $n\in {\cal N}$ appears on an active path
sharing a node with the equivalence class $c\in {\cal C}$.
Equation~\ref{eq6} defines an active path that shares at least
one node with two classes. Finally, equation~\ref{eq7} defines the
load of a node as the sum of the flows associated to active paths
passing through node $n$. \\

%\vspace{-.08in}
\noindent\textbf{Objective}\\
\begin{equation}\label{obj1}
  \min \left(
    \begin{array}{lll}      
      \alpha_0 & \sum_{p, T}{len(p) * flow_{p, T}} &+ \\
      \alpha_1 & \sum_{(c_1,c_2)\in {\cal R}, n\in{\cal N}}{(share_{c_1, c_2, n} - 1)} &+ \\
      \alpha_2 & \sum_{n \in {\cal N}}{(load_n)^2} &
    \end{array}
  \right)
\end{equation}
The objective function~\ref{obj1} in this model is a weighted sum of three
terms. The first term captures the total flows which are penalized by
the length of the path used to dispatch those flows (such policies are
codified in OSFP~\cite{moy1998ospf} and BGP practice~\cite{gill2013survey}).
The second term gives a unit credit each time equivalence classes on the
segregation list ${\cal R}$ do not share a node.  (Due to this term, the objective value of the 
final solution may change between iterations of the functional layer.)
The third and final term contribute to a
bias towards solutions that achieve load balancing thanks to the
quadratic component which heavily penalizes nodes with large loads.
% \ben{is the only quadratic part of problem, or does the multiplication
%   of equiv count?  should we draw this out?}
% \ldm{This is the only part. The conjunction of $equiv$ can be
%   implemented as a linear inequality. I had a conversation with Tim
%   about this pointing out that we could have an iterative scheme
%   that progressively tighten the bound on variables capturing the
%   ``distance to mean'' and that would make the problem linear. I
%   suggested (because of the time we have left) to leave this for later.}

\noindent\textbf{Solving the Functional Model}
The functional model
starts with empty sets ${\cal C}$ and ${\cal R}$ which are augmented
with each iteration of the framework.  New
sets of nodes are added to  ${\cal C}$ and new segregation rules are
added to ${\cal R}$ (by the security layer).
In the current implementation, least cost paths between pairs of nodes $s,t$ are not
generated ``on demand''. Instead, the generation is limited to the
first best $k$ such paths, for increasing values of $k$. This process
will ultimately be improved to use column generation techniques~\cite{Dantzig:1960}. 

%% --------------------------------------------------------------------------------
%% --------------------------------------------------------------------------------
%% --------------------------------------------------------------------------------

\subsection{Risk Calculation}
\label{sec:risk}
After the functional layer finds an optimal solution, it passes this solution to the 
risk calculation procedure. This input is the set of active paths. This module calculates the 
effective risk to the network for each path and traffic type.  

\noindent\textbf{Inputs}\\
%risk info
$risk(n, T) : {\cal N} \times {\cal T} \rightarrow \mathbb{R}$ -- the risk 
inherit to network device $n$ for traffic of type $T$ ($risk(n,T) \geq1$)\\
% dist k nodes
$d_k(n) : {\cal N} \rightarrow 2^{\cal N}$ -- the set of nodes at a distance at most $k$ from $n$ 
in the logical topology\\
% active flows

\vspace{-.05in}
\noindent\textbf{Calculation}\\
Given an \emph{active} path $p \in {\cal P}$ with source $s$ and sink $t$, the calculation proceeds
by partitioning the set of nodes of the path into three segments: the
nodes ``close'' to the source $s$, ``close'' to the sink $t$ and the
nodes ``in between''. Closeness is characterized by the function $d_k$
and is meant to capture any connected node over the logical topology
which sits no more that $k$ hops away. Given this partition,
$flowRisk(p,T)$ is:
\[
  \begin{array}{lcl}
    flowRisk(p, T) &=& \sum_{i \in d_2(s) \cup d_2(t)} risk(i,T)^2 + \\
    && \sum_{i \in N(p) \setminus \left(d_2(s) \cup d_2(t)\right)} risk(i,T)^2
  \end{array}
\]
We use $k=2$ to model 
nodes on the same LAN.  
The rationale is to impart to source $s$ and sink $t$ risk resulting
from \emph{lateral
  movement} of attacks. All other nodes contribute to the overall path
risk in proportion to the square of their own risks.  We expect in most networks
for $d_2(s)$ (and $d_2(t)$) to include nodes not directly on the path (like nodes 
on the same LAN).
%Clearly $d_2(s)
%\cap N(p) \neq \emptyset$  and $d_2(t) \cap N(p) \neq \emptyset$ since
%$d_2$ includes neighbors that are \emph{not} on $p$. 
The input path risk calculation
$flowRisk(p,T)$ is modular and can be
augmented using other risk calculation methods.
%% --------------------------------------------------------------------------------
%% --------------------------------------------------------------------------------
%% --------------------------------------------------------------------------------

\subsection{Security Layer}
\label{sec:sec}
The mathematical optimization model in the security layer is a
 mixed integer programming model. We similarly present
the inputs, variables, constraints, and objective for the security
layer.  Its inputs are given
below. Also note that all the variables from the functional model are
\emph{constants}.

\noindent\textbf{Inputs}\\
$mem(n) : {\cal N} \rightarrow \mathbb{R}$ -- the memory resources of SDN device $n$\\
$fwCost(T) : {\cal T} \rightarrow \mathbb{R}$ -- the memory footprint for a firewall blocking traffic type $T$\\
$piCost$ -- the memory footprint for a packet inspection post\\
%$len(p): {\cal P} \rightarrow {\mathbb{Z}}$ -- the length of path $p$\\
$fwComp$ -- the complexity footprint for adding a firewall\\
$piComp$ -- the complexity footprint for adding a packet inspection post to the network\\
$penalty(p, T) : {\cal P} \times {\cal T} \rightarrow {\mathbb{R}}$ -- the penalty for blocking 
a unit of flow of type $T$ along path $p$\\
$rank(n, p) : {\cal N} \times {\cal P} \rightarrow {\mathbb{Z}}$ -- the position of node $n$ in path $p$\\
$flowRisk(p, T) : {\cal P} \times {\cal T} \rightarrow
\mathbb{R}_{\geq 0}$ -- above risk calculation

\noindent\textbf{Variables}\\
$fw_{n, T} \in \{0,1\}$ -- does a firewall block traffic type $T$ at $n$\\
$pi_{n} \in \{0,1\}$ -- is there packet inspection at network device $n$\\
%artifical variables
$fwOR_{n ,T} \in \{0,1\}$ -- is there a \emph{block everything} or
\emph{block traffic of type $T$} firewall at network device $n$\\ 
$fwOP_{p, T} \in \{0,1\}$ -- is there a firewall  on path $p$\\
%$piOP_{p} \in \{0,1\}$ -- is there packet inspection on path $p$\\
$rf_{p, T} \in [0,1]$ -- risk factor for path $p\in
P(s,t)$ servicing flow $(s,t,T)\in {\cal F}$\\ 
$RMfw_{p, n, T} \in [0,1]$ -- used in the riskFactor calculation\\
$RMpi_{p, n, T} \in [0,1]$ -- used in the riskFactor calculation

\noindent\textbf{Constraints}\\
\vspace{-.05in}
\begin{equation}
  \label{m2:eq8} % riskPI
  fwOR_{n,T} = fw_{n, T} \lor fw_{n,*}, \forall n \in {\cal N}, T \in {\cal T}
\end{equation}
\begin{equation}
  \label{m2:eq1} % device memory
  \sum_{T \in{\cal T} \cup \{*\}}{fwCost_T \cdot fw_{n, T}} + {piCost \cdot pi_n} \leq
  mem_n, \forall n \in {\cal N}
\end{equation}
%
%\begin{equation}
%  \label{m2:eq2} % piOnPath
%  piOP_p = \!\!\bigvee_{n \in N(p)} (pi_n), \forall T\in {\cal T},p \in {\cal P} : active_{p,T}
%\end{equation}
%
\begin{equation}
  \label{m2:eq3} % piOnPath
  fwOP_{p, T} \!=\!\!\!\bigvee_{n \in N(p)} \!\!(fwOR_{n, T}), \forall T\in {\cal T},p \in {\cal P} : active_{p,T}
\end{equation}
\vspace{.05in}
%\begin{equation}
%  \label{m2:eq4} % piOnPath
%fwOP_{p, T} + piOP_p \geq 1,\forall T\in{\cal T},p \in {\cal P} : active_{p,T}
%\end{equation}
%
\begin{equation}
  \label{m2:eq5} % riskFW
  \begin{aligned}
    %&RMfw_{p,n,T} = 1 - (1 - \frac{rank(n,p)}{len(p)}) \cdot fwOR_{n,T} \\
    &RMfw_{p,n,T} = 1 - (.5)^{rank(n,p)} \cdot fwOR_{n,T}, \\
    &\forall p \in P(s,t),n \in N(p),(s,t,T) \in {\cal F} 
  \end{aligned}
\end{equation}
\begin{equation}
  \label{m2:eq6} % riskPI
  \begin{aligned}
    &RMpi_{p, n, T} = 1 - 0.1 \cdot (.5)^{rank(n,p)} \cdot pi_n,\\
    &\forall p \in P(s,t),n \in N(p),(s,t,T) \in {\cal F} 
  \end{aligned}
\end{equation}
\begin{equation}
  \label{m2:eq7} % riskPI
%rf_{p, T} = \min_{n \in N(p)} \min(RMfw_{p, n, T},RMpi_{p, n, T})
  \begin{aligned}
    rf_{p, T} = \min \ &\bigcup_{n \in N(p)} \{RMfw_{p, n, T}, RMpi_{p, n, T}\}, \\
    &\forall T \in {\cal T}, p \in {\cal P} : active_{p, T}
  \end{aligned}
\end{equation}
\noindent
Equation~\ref{m2:eq8} is used to define the presence of a
firewall that will block traffic of type $T$ at a node $n$.
 Equation~\ref{m2:eq1} ensures that the memory footprint in
SDN node $n$ for the deployment of the firewall and the packet
inspection logic does not exceed the device memory.
%
%
%Equation~\ref{m2:eq2} links the packet inspection on a path with
%the presence of packet inspection on any node along the active
%path.
Equation~\ref{m2:eq3} links the presence of a firewall that will block 
traffic of type $T$ on a path with 
the presence of a firewall that will block traffic of type $T$
on any node along the active
path.
%Equation~\ref{m2:eq4} ensures that a firewall or
%a packet inspection is deployed somewhere on an active
%path.

Equation~\ref{m2:eq5} defines the minimum risk factor associated
to a firewall. The earlier on the path the firewall is
deployed, the lower the risk. Equation~\ref{m2:eq6} similarly defines
the minimal risk. Equation~\ref{m2:eq7} defines the composite risk
factor.

\vspace{.05in}
\noindent\textbf{Objective}\\
\begin{equation}\label{obj2}
  \min \left(\!\!
    \begin{array}{lll}
      %\beta_0 \left(\sum_{n, T}{fwCost_T \cdot fw_{n, T}} +
      %\sum_{n}{piCost \cdot pi_n}\right) + \\
      \beta_0 \left(\sum_{n, T}{fwComp \cdot fw_{n, T}} +
      \sum_{n}{piComp \cdot pi_n}\right) + \\
      \beta_1 \sum_{n}{load_n \cdot pi_n} + \\
      \beta_2 \sum_{p, T}{penalty(p, T) \cdot flow_{p, T} \cdot fwOP_{p, T}}+\\
      \beta_3 \sum_{p, T}{flowRisk_{p, T} \cdot rf_{p, T}}
    \end{array}\!\!
  \right)
\end{equation}
The objective function defined in equation~\ref{obj2} is a weighted
sum of four distinct terms that focus on minimizing the network complexity based on security 
resources deployed, the load induced by inspection posts, the penalties incurred from dropping 
desirable flows due to firewall placement and finally the residual risk. This
model is a classic mixed integer programming
formulation.

%% --------------------------------------------------------------------------------
%% --------------------------------------------------------------------------------
%% --------------------------------------------------------------------------------

\subsection{Result Analysis}
\label{sec:analysis}
The result analysis module tries to generate cuts 
for the functional layer with the goal of improving both functionality and security. 
To generate cuts, this module will form equivalence classes of network nodes and pass back 
certain pairs of these classes, one at a time, to the functional layer. Each pair of classes 
describes a segregation rule, or a cut, to which to functional layer will adhere to as much as possible.
 
After the functional and security layers are re-optimized using the most recent cut, 
the result analysis module determines whether the cut was beneficial or harmful based on the 
objectives of each layer. If the cut is deemed to have been beneficial, we permanently keep it as a constraint, 
repopulate the cut queue, and continue the process.

If the cut is deemed to have been harmful, it is removed from the functional 
layer's constraint pool. Then the next cut in the queue will be passed back to the functional layer.
If the cut queue is empty, the feedback mechanism terminates and we output the best solution found.

We note that since this process only provides pairs of nodes it is a heuristic, it may be necessary for many nodes to simultaneously be separated to arrive at a global optimum.  This mechanism performed well in our experiments.

%% --------------------------------------------------------------------------------
%% --------------------------------------------------------------------------------
%% --------------------------------------------------------------------------------

\subsection{Layer Coordination}
\label{sec:bender}
It is valuable to review how the layers coordinate. 
The functional layer sends
to the security layer a set of paths that implements the routing
within the network to serve the specified flows while satisfying a set
of segregation requirements. The security layer first computes
\emph{risks} for these paths (in polynomial time) based on its
knowledge of the traffic. The paths, their risk and the security model
are then tasked with deploying packet inspection apparatus as well as
firewalls within that logical topology to monitor the traffic and
block threats (risky traffic). Once the security model is solved to
optimality, an analysis can determine whether the proposed logical
topology is beneficial or not (w.r.t. its objective) and even 
suggest further equivalence classes for network
nodes as well as segregation rules to be sent back to the
functional layer for another iteration.
Fundamentally, the coordination signal boils down to additional
equivalence classes to group nodes together with segregation rules to
separate paths that include network nodes in ``antagonistic''
equivalence classes. 

\vspace{-.05in}
\subsection{Outputs}
\label{sec:outputs}

When the set of potential cuts is empty, the proposed configuration
can be parsed and translated into SDN language fragments to be
deployed on the network devices in order to obtain the desired logical
network topology put forth by our framework.

%%% Local Variables:
%%% mode: latex
%%% TeX-master: "main"
%%% End:

%!TEX root = main.tex

\section{Experimental Setup}
\label{sec:setup} 
A fundamental component of our work is the
separation of the physical and logical networks.  Our framework has
potential in applications where many different logical topologies are
possible from a single physical topology. Physical topology is an input to our framework and the empirical
evaluation is based on a popular topology: 
Fat-Tree~\cite{Al-Fares:2008:SCD:1402958.1402967}.

\begin{figure}[t]
  \begin{center}
%    \vspace{.05in}
    \includegraphics[width=0.47\textwidth]{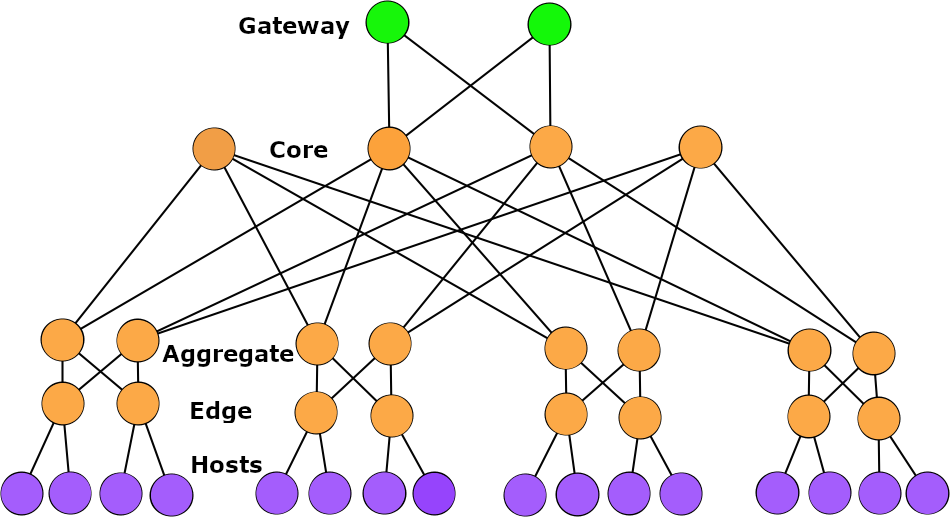}
%    \vspace{-4mm}
\end{center}
\caption{\small Order 4 Fat-tree with 2 gateway switches at the top and 2 hosts per edge switch.}
\label{fig:fat-tree}
\end{figure}
 
The instance of Fat-Tree we use is shown in Figure~\ref{fig:fat-tree}.
The updated design avoids bottlenecks through multiple equal capacity
links between layers.   
This design uses four layers of switches: gateway, core, aggregate and edge
with hosts connected at the bottom. 

Within our sample
network, we consider having two main types of devices:
switches/routers and hosts.  In order to model traffic between
internal and external entities we utilize two gateway switches which
represent the boundary of our network. For generality we consider
 two traffic types A and B which could represent any
types of traffic such as web and storage.    
We also classify traffic as internal and external, with external
traffic traversing one of the gateways.  We allow only half of our
hosts to communicate with external sources by allowing them to connect
to one of the two gateways. Further, all hosts are involved in
internal communications. In our network we have 16 hosts and we
generated 60 flows, 44 of them being internal and 16 being
external. 

Additionally, our setup simulates an emergent
vulnerability/active attack.  We select two hosts that are highly vulnerable to, or being
targeted on, a specific type of traffic, resulting in a significant
increase in their risk for the corresponding type of traffic. In particular, this 
could represent at DDoS attack on these two hosts.

We run multiple experiments providing the framework
increasing numbers of starting paths  between source and destination (from the priming procedure) to
determine the impact on the solution quality. 

Our implementation was built in Python 3.6 using the Gurobi 8.1 optimization library~\cite{optimization2014inc}.
The experiments were run on a machine running Ubuntu 18.04 and
equipped with an Intel Core i9-8950HK processor operating at 2.90 GHz
with a 12 MB Cache and 32 GB of physical memory.

%%% Local Variables:
%%% mode: latex
%%% TeX-master: "main"
%%% End:

%!TEX root = main.tex
\section{Evaluation/Results}
\label{sec:eval}

\begin{figure*}[t]
  \begin{center}
    %\vspace{-5mm}
    \includegraphics[width=\textwidth]{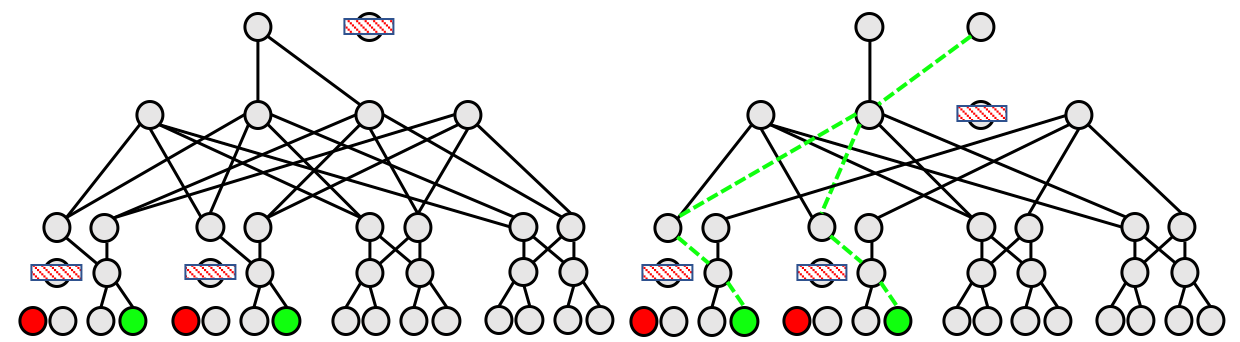}
    \vspace{-4mm}
\end{center}
\caption{{\small Illustration of the Fat-Tree network after the first pass through both layers of the framework (left) and the final solution (right).  Note: Firewalls are depicted with rectangles, red nodes represent high risk nodes, green nodes represent nodes that are initially blocked and recovered in the final solution, utilizing the updated routes shown in dashed lines.  The modification of the logical topology allows for more intelligent firewall placement balancing both functionality and security. }}
\label{fig:results}
\end{figure*}

%\subsection{Fat-Tree}
We begin with the model's resolution on the above scenario using the 10
shortest paths per source and destination pair to prime the optimization. We 
assume equal demand for each of the 60 flows and two high risk hosts
(in red in Figure~\ref{fig:results}).  In the first iteration, 
the functional layer of the  framework generates a candidate topology 
and passes this
solution to the security layer. The security layer then 
calculates the network risk and deploys firewalls (represented with
rectangles). 

In the first iteration, the gateway on the right serves
both low risk (shown in green) and high risk hosts (shown in red). 
Deploying a  firewall on this gateway significantly reduces the
network risk and is the output of the security layer.
Importantly, this results in collateral damage as flows to low risk hosts are blocked.
In total, the first iteration through the framework deploys 3 firewalls
which block 12 flows, 8 of which are high risk. 

Iteratively, the security layer 
proposes separation of these collateral nodes from the high-risk nodes.
The solution of the last
framework iteration (493 candidates cuts are proposed, 40 prove beneficial) is shown on the right.  
This configuration routes all high risk flows
through one core switch where a firewall is now deployed.  Meanwhile,
all the low risk flows access the gateways through a separate core
switch. 

\begin{table*}
  \begin{center}
  \csvreader[tabular=| l | r | r | r | r | r | r | r,
table head=\hline & 10 paths & 20 paths & 30 paths & 40 paths & 50 paths & 100 paths\\\hline\hline,
late after line=\\\hline]%
{pathTable.csv}{0Paths=\pathzero,10Paths=\pathten,20Paths=\pathtwen,30Paths=\paththir,40Paths=\pathfour,50Paths=\pathfif,100Paths=\pathhun}{\pathzero & \pathten & \pathtwen & \paththir & \pathfour & \pathfif& \pathhun}%
\end{center}
\caption{Experimental results from applying the \dsdn{} framework to an order 4 Fat-tree. Each column refers to a separate experiment where the number of paths per source-destination pair given to the framework were varied. 
Note that the functional objective values in this table are calculated without the cut reward, the second term in Equation~\ref{obj1}, in order to facilitate comparisons across columns.}
\label{tab:table}
\end{table*}

Overall we conducted six experiments modifying only the number of paths ($\{10, 20, 30, 40, 50, 100\}$) 
being used to prime the functional layer for each source-destination pair. Complete experimental 
results are shown in Table~\ref{tab:table}. Ultimately, this process will be
dynamic and use column-generation. 
A few observations are in order:
\begin{itemize}
\item The overall objective reflects both the functional and security
  layers. The other objective rows refer to each layer objective
  individually.  The network risk rows quantify the risks and their
  change as the optimization proceeds. For instance, for the 10 paths
  benchmark, the risk degrades from 10425 to 11646 or 11.7\% as a
  result of supporting an additional 4 good flows.
\item The ``nodes explored'' rows indicate of the size of
  the branch and bound tree and remains quite modest throughout. 
 \item  Within each experiment we observe a meaningful search,
  as seen by numerous cuts sent back to the functional layer, to 
  segregate high and low risk flows.
\item All runs blocked all flows that contain a high risk host while
  preserving the low risk flows. All experiments delivered final
  configurations that preserved the same low-risk flows. We therefore
  hypothesize that a column-generation would quickly settle down and
  prove that no additional path can improve the quality of the
  solution. It is nonetheless interesting that adding more paths does
  not negatively impact the overall runtime.
\item The objective functions of the functional and security layers
  use ``scores'' meant to ease the interplay between the two. Yet, it
  is wise to consult the \emph{raw} properties of the solutions to
  appreciate the impact of the optimization. In particular, the number
  of flows blocked and the network risks. What is readily apparent is
  that improving functionality induces a slight degradation in the
  network risks, underlying the conflicting nature of the two
  objectives. The individual objective scores while moving in the
  correct direction are not to be viewed as stand alone metrics to
  determine solution quality but rather inter layer communications
  indicating improvement or decline from a functional or security
  perspective.
\item The objective scores vary across our experiments due to the stochasticity 
 introduced by our heuristic-driven feedback module (see Section~\ref{sec:analysis} for discussion). 
 For instance, the functional objecive in the 30 path experiment is slightly worse than it is 
 in other runs, but this difference does not impact the number of serviced flows
 in the final configuration.
\item The variance in time, iterations and number of cuts produced by
  each experiment is due to symmetries in the formulation.  Solutions
  that are symmetric in the functional layer may not be symmetric in
  the security layer and induce slightly different solutions
  there. This is especially true for a
  Fat-tree network due to its built in redundancy/symmetry. 
\item Beneficial cuts reflects the number of segregation proposals from the
  security layers that are adopted by the functional layer (these cuts
  remove the current best feasible solution). harmful cuts are segregation
  proposals that do not ``cut'' the current best feasible solution or
  worsen the functional solution. 
\end{itemize}

\section{Conclusion}
\label{sec:conclusion}

Our framework is portable with respect to network risk assessment. Since the risk calculation/analysis 
is decoupled from the optimization model, the framework can be combined with any procedure 
that calculates risk on a per path basis. Along with this procedure, the other requirements for implementing a different risk mechanism:
\begin{itemize}
\item A way of evaluating how risk changes due to the deployment of network defenses
\item The ability to propose candidate cuts that can be passed to the functional layer. 
\end{itemize}

Our results show it is possible to effectively, automatically, and quickly  find a network configuration that  meets multiple conflicting properties. 
Our framework is modular, enabling integration of new desired properties.  \dsdn will allow network administrators to effectively prioritize and choose their desired properties.  The efficiency of \dsdn{} is enabled by the feedback/interplay between the functional and security optimization layers.

\bibliographystyle{IEEEtran}
\IEEEtriggeratref{32}
{%\footnotesize
 \bibliography{abbrev3,biblio,wireless-management}

% Generated by IEEEtran.bst, version: 1.14 (2015/08/26)
\begin{thebibliography}{10}
\providecommand{\url}[1]{#1}
\csname url@samestyle\endcsname
\providecommand{\newblock}{\relax}
\providecommand{\bibinfo}[2]{#2}
\providecommand{\BIBentrySTDinterwordspacing}{\spaceskip=0pt\relax}
\providecommand{\BIBentryALTinterwordstretchfactor}{4}
\providecommand{\BIBentryALTinterwordspacing}{\spaceskip=\fontdimen2\font plus
\BIBentryALTinterwordstretchfactor\fontdimen3\font minus
  \fontdimen4\font\relax}
\providecommand{\BIBforeignlanguage}[2]{{%
\expandafter\ifx\csname l@#1\endcsname\relax
\typeout{** WARNING: IEEEtran.bst: No hyphenation pattern has been}%
\typeout{** loaded for the language `#1'. Using the pattern for}%
\typeout{** the default language instead.}%
\else
\language=\csname l@#1\endcsname
\fi
#2}}
\providecommand{\BIBdecl}{\relax}
\BIBdecl

\bibitem{khurshid2012veriflow}
A.~Khurshid, W.~Zhou, M.~Caesar, and P.~Godfrey, ``Veriflow: Verifying
  network-wide invariants in real time,'' in \emph{Proceedings of the first
  workshop on Hot topics in software defined networks}.\hskip 1em plus 0.5em
  minus 0.4em\relax ACM, 2012, pp. 49--54.

\bibitem{neves2016selfnet}
P.~Neves, R.~Cal{\'e}, M.~R. Costa, C.~Parada, B.~Parreira, J.~Alcaraz-Calero,
  Q.~Wang, J.~Nightingale, E.~Chirivella-Perez, W.~Jiang \emph{et~al.}, ``The
  {SELFNET} approach for autonomic management in an {NFV/SDN} networking
  paradigm,'' \emph{International Journal of Distributed Sensor Networks},
  vol.~12, no.~2, p. 2897479, 2016.

\bibitem{Schneier2}
\BIBentryALTinterwordspacing
B.~Schneier, ``Attack trees,'' {Blog}, 1999. [Online]. Available:
  \url{http://tnlandforms.us/cs594-cns96/attacktrees.pdf}
\BIBentrySTDinterwordspacing

\bibitem{yu2018deploying}
R.~Yu, G.~Xue, V.~T. Kilari, and X.~Zhang, ``Deploying robust security in
  internet of things,'' in \emph{IEEE Conference on Computer and Network
  Security}, 2018.

\bibitem{wang2011openflow}
R.~Wang, D.~Butnariu, J.~Rexford \emph{et~al.}, ``Openflow-based server load
  balancing gone wild.'' \emph{Hot-ICE}, vol.~11, pp. 12--12, 2011.

\bibitem{skowyra2014verification}
R.~Skowyra, A.~Lapets, A.~Bestavros, and A.~Kfoury, ``A verification platform
  for {SDN}-enabled applications,'' in \emph{Cloud Engineering (IC2E), 2014
  IEEE International Conference on}.\hskip 1em plus 0.5em minus 0.4em\relax
  IEEE, 2014, pp. 337--342.

\bibitem{Bixby2000}
R.~Bixby, M.~Fenelon, Z.~Gu, E.~Rothberg, and R.~Wunderling, \emph{{System
  Modelling and Optimization: Methods, Theory, and Applications}}.\hskip 1em
  plus 0.5em minus 0.4em\relax Kluwer Academic Publishers, 2000, ch. MIP:
  Theory and practice -- closing the gap, pp. 19--49.

\bibitem{ASS2015}
B.~Fourer, ``Amazing solver speedups,'' online, 2015,
  \url{http://bob4er.blogspot.com/2015/05/amazing-solver-speedups.html}.

\bibitem{Benders1962}
J.~F. Benders, ``{Partitioning procedures for solving mixed-variables
  programming problems},'' \emph{Numerische Mathematik}, vol.~4, no.~1, pp.
  238--252, 1962.

\bibitem{Hooker95}
J.~Hooker, ``Logic-based {B}enders decomposition,'' \emph{Mathematical
  Programming}, vol.~96, p. 2003, 1995.

\bibitem{HookerBook}
------, \emph{{Logic-Based Methods for Optimization: Combining Optimization and
  Constraint Satisfaction}}.\hskip 1em plus 0.5em minus 0.4em\relax John Wiley
  and Sons, 2000.

\bibitem{Codato2006}
\BIBentryALTinterwordspacing
G.~Codato and M.~Fischetti, ``Combinatorial {B}enders' cuts for mixed-integer
  linear programming,'' \emph{Operations Research}, vol.~54, no.~4, pp.
  756--766, 2006. [Online]. Available:
  \url{http://dx.doi.org/10.1287/opre.1060.0286}
\BIBentrySTDinterwordspacing

\bibitem{barnhart1998branch}
C.~Barnhart, E.~L. Johnson, G.~L. Nemhauser, M.~W. Savelsbergh, and P.~H.
  Vance, ``Branch-and-price: Column generation for solving huge integer
  programs,'' \emph{Operations research}, vol.~46, no.~3, pp. 316--329, 1998.

\bibitem{Hijazi:2015}
\BIBentryALTinterwordspacing
H.~Hijazi, T.~W.~K. Mak, and P.~Van~Hentenryck, ``Power system restoration with
  transient stability,'' in \emph{Proceedings of the Twenty-Ninth AAAI
  Conference on Artificial Intelligence}, ser. AAAI'15.\hskip 1em plus 0.5em
  minus 0.4em\relax AAAI Press, 2015, pp. 658--664. [Online]. Available:
  \url{http://dl.acm.org/citation.cfm?id=2887007.2887099}
\BIBentrySTDinterwordspacing

\bibitem{Lam:2016}
\BIBentryALTinterwordspacing
E.~Lam and P.~V. Hentenryck, ``A branch-and-price-and-check model for the
  vehicle routing problem with location congestion,'' \emph{Constraints},
  vol.~21, no.~3, pp. 394--412, Jul. 2016. [Online]. Available:
  \url{http://dx.doi.org/10.1007/s10601-016-9241-2}
\BIBentrySTDinterwordspacing

\bibitem{Dantzig:1960}
\BIBentryALTinterwordspacing
G.~B. Dantzig and P.~Wolfe, ``Decomposition principle for linear programs,''
  \emph{Oper. Res.}, vol.~8, no.~1, pp. 101--111, Feb. 1960. [Online].
  Available: \url{http://dx.doi.org/10.1287/opre.8.1.101}
\BIBentrySTDinterwordspacing

\bibitem{Shaw98}
P.~Shaw, ``{Using Constraint Programming and Local Search Methods to Solve
  Vehicle Routing Problems},'' in \emph{{Proceedings of Fourth International
  Conference on the Principles and Practice of Constraint Programming
  (CP'98)}}.\hskip 1em plus 0.5em minus 0.4em\relax Springer Verlag, October
  1998, pp. 417--431.

\bibitem{skottler2018}
\BIBentryALTinterwordspacing
S.~Kottler. (2018, March) February 28th ddos incident report. [Online].
  Available: \url{https://githubengineering.com/ddos-incident-report/}
\BIBentrySTDinterwordspacing

\bibitem{Ioannidis01pushback:router-based}
J.~Ioannidis and S.~M. Bellovin, ``Pushback: Router-based defense against
  {DDoS} attacks,'' 2001.

\bibitem{ioannidis2002implementing}
------, ``Implementing pushback: Router-based defense against {DDoS} attacks.''
  in \emph{NDSS}, vol.~2, 2002.

\bibitem{mirkovic2004taxonomy}
J.~Mirkovic and P.~Reiher, ``A taxonomy of {DDoS} attack and {DDoS} defense
  mechanisms,'' \emph{ACM SIGCOMM Computer Communication Review}, vol.~34,
  no.~2, pp. 39--53, 2004.

\bibitem{peng2007survey}
T.~Peng, C.~Leckie, and K.~Ramamohanarao, ``Survey of network-based defense
  mechanisms countering the {DoS} and {DDoS} problems,'' \emph{ACM Computing
  Surveys (CSUR)}, vol.~39, no.~1, p.~3, 2007.

\bibitem{zargar2013survey}
S.~T. Zargar, J.~Joshi, and D.~Tipper, ``A survey of defense mechanisms against
  distributed denial of service ({DDoS}) flooding attacks,'' \emph{IEEE
  communications surveys \& tutorials}, vol.~15, no.~4, pp. 2046--2069, 2013.

\bibitem{fayaz2015bohatei}
S.~K. Fayaz, Y.~Tobioka, V.~Sekar, and M.~Bailey, ``Bohatei: Flexible and
  elastic {DDoS} defense.'' in \emph{USENIX Security Symposium}, 2015, pp.
  817--832.

\bibitem{marczak2015china}
B.~Marczak, N.~Weaver, J.~Dalek, R.~Ensafi, D.~Fifield, S.~McKune, A.~Rey,
  J.~Scott-Railton, R.~Deibert, and V.~Paxson, ``China’s great cannon,''
  \emph{Citizen Lab}, vol.~10, 2015.

\bibitem{reich2013modular}
J.~Reich, C.~Monsanto, N.~Foster, J.~Rexford, and D.~Walker, ``Modular {SDN}
  programming with {P}yretic,'' \emph{Technical Reprot of USENIX}, 2013.

\bibitem{Benson:2010:NTC:1879141.1879175}
\BIBentryALTinterwordspacing
T.~Benson, A.~Akella, and D.~A. Maltz, ``Network traffic characteristics of
  data centers in the wild,'' in \emph{Proceedings of the 10th ACM SIGCOMM
  Conference on Internet Measurement}, ser. IMC '10.\hskip 1em plus 0.5em minus
  0.4em\relax New York, NY, USA: ACM, 2010, pp. 267--280. [Online]. Available:
  \url{http://doi.acm.org/10.1145/1879141.1879175}
\BIBentrySTDinterwordspacing

\bibitem{kreutz2015software}
D.~Kreutz, F.~M. Ramos, P.~E. Verissimo, C.~E. Rothenberg, S.~Azodolmolky, and
  S.~Uhlig, ``Software-defined networking: A comprehensive survey,''
  \emph{Proceedings of the IEEE}, vol. 103, no.~1, pp. 14--76, 2015.

\bibitem{mckeown2008openflow}
N.~McKeown, T.~Anderson, H.~Balakrishnan, G.~Parulkar, L.~Peterson, J.~Rexford,
  S.~Shenker, and J.~Turner, ``Openflow: enabling innovation in campus
  networks,'' \emph{ACM SIGCOMM Computer Communication Review}, vol.~38, no.~2,
  pp. 69--74, 2008.

\bibitem{foster2011frenetic}
N.~Foster, R.~Harrison, M.~J. Freedman, C.~Monsanto, J.~Rexford, A.~Story, and
  D.~Walker, ``Frenetic: A network programming language,'' \emph{ACM Sigplan
  Notices}, vol.~46, no.~9, pp. 279--291, 2011.

\bibitem{kim2015kinetic}
H.~Kim, J.~Reich, A.~Gupta, M.~Shahbaz, N.~Feamster, and R.~J. Clark,
  ``Kinetic: Verifiable dynamic network control.'' in \emph{NSDI}, 2015, pp.
  59--72.

\bibitem{beckett2017network}
R.~Beckett, R.~Mahajan, T.~Millstein, J.~Padhye, and D.~Walker, ``Network
  configuration synthesis with abstract topologies,'' in \emph{Proceedings of
  the 38th ACM SIGPLAN Conference on Programming Language Design and
  Implementation}.\hskip 1em plus 0.5em minus 0.4em\relax ACM, 2017, pp.
  437--451.

\bibitem{website:CVE}
\BIBentryALTinterwordspacing
M.~Corp, ``Common vulnerabilities and exposures,'' December 2018. [Online].
  Available: \url{https://cve.mitre.org}
\BIBentrySTDinterwordspacing

\bibitem{website:NVD}
\BIBentryALTinterwordspacing
NIST, ``National vulnerability database,'' December 2018. [Online]. Available:
  \url{https://nvd.nist.gov}
\BIBentrySTDinterwordspacing

\bibitem{stoneburner2002sp}
G.~Stoneburner, A.~Y. Goguen, and A.~Feringa, ``{SP} 800-30. risk management
  guide for information technology systems,'' 2002.

\bibitem{jansen2010directions}
W.~Jansen, \emph{Directions in security metrics research}.\hskip 1em plus 0.5em
  minus 0.4em\relax Diane Publishing, 2010.

\bibitem{stolfo2011measuring}
S.~Stolfo, S.~M. Bellovin, and D.~Evans, ``Measuring security,'' \emph{IEEE
  Security \& Privacy}, vol.~9, no.~3, pp. 60--65, 2011.

\bibitem{lippmann2012continuous}
R.~Lippmann, J.~Riordan, T.~Yu, and K.~Watson, ``Continuous security metrics
  for prevalent network threats: introduction and first four metrics,''
  Massachusetts Inst of Tech Lexington Lincoln Lab, Tech. Rep., 2012.

\bibitem{cherdantseva2016review}
Y.~Cherdantseva, P.~Burnap, A.~Blyth, P.~Eden, K.~Jones, H.~Soulsby, and
  K.~Stoddart, ``A review of cyber security risk assessment methods for scada
  systems,'' \emph{Computers \& security}, vol.~56, pp. 1--27, 2016.

\bibitem{lippmann2016threat}
R.~P. Lippmann and J.~F. Riordan, ``Threat-based risk assessment for enterprise
  networks,'' \emph{Lincoln Laboratory Journal}, vol.~22, no.~1, pp. 33--45,
  2016.

\bibitem{ingols2006practical}
K.~Ingols, R.~Lippmann, and K.~Piwowarski, ``Practical attack graph generation
  for network defense,'' in \emph{Computer Security Applications Conference,
  2006. ACSAC'06. 22nd Annual}.\hskip 1em plus 0.5em minus 0.4em\relax IEEE,
  2006, pp. 121--130.

\bibitem{kaynar2016taxonomy}
K.~Kaynar, ``A taxonomy for attack graph generation and usage in network
  security,'' \emph{Journal of Information Security and Applications}, vol.~29,
  pp. 27--56, 2016.

\bibitem{zhang2013sat}
S.~Zhang and S.~Malik, ``{SAT} based verification of network data planes,'' in
  \emph{International Symposium on Automated Technology for Verification and
  Analysis}.\hskip 1em plus 0.5em minus 0.4em\relax Springer, 2013, pp.
  496--505.

\bibitem{beckett2017general}
R.~Beckett, A.~Gupta, R.~Mahajan, and D.~Walker, ``A general approach to
  network configuration verification,'' in \emph{Proceedings of the Conference
  of the ACM Special Interest Group on Data Communication}.\hskip 1em plus
  0.5em minus 0.4em\relax ACM, 2017, pp. 155--168.

\bibitem{Rossi:2006:HCP:1207782}
F.~Rossi, P.~v. Beek, and T.~Walsh, \emph{Handbook of Constraint Programming
  (Foundations of Artificial Intelligence)}.\hskip 1em plus 0.5em minus
  0.4em\relax New York, NY, USA: Elsevier Science Inc., 2006.

\bibitem{Baptiste2001}
P.~Baptiste, C.~Le~Pape, and W.~Nuijten, \emph{{Constraint-Based
  Scheduling}}.\hskip 1em plus 0.5em minus 0.4em\relax Kluwer Academic
  Publishers, 2001.

\bibitem{DBLP:conf/pscc/NagarajanYBHB16}
H.~Nagarajan, E.~Yamangil, R.~Bent, P.~V. Hentenryck, and S.~Backhaus,
  ``Optimal resilient transmission grid design,'' in \emph{{PSCC}}.\hskip 1em
  plus 0.5em minus 0.4em\relax {IEEE}, 2016, pp. 1--7.

\bibitem{Byeon18}
G.~{Byeon}, P.~{Van Hentenryck}, R.~{Bent}, and H.~{Nagarajan},
  ``{Communication-Constrained Expansion Planning for Resilient Distribution
  Systems},'' \emph{ArXiv e-prints}, Jan. 2018.

\bibitem{DBLP:conf/cds/CoattaN92}
T.~Coatta and G.~W. Neufeld, ``Configuration management via constraint
  programming,'' in \emph{{CDS}}.\hskip 1em plus 0.5em minus 0.4em\relax
  {IEEE}, 1992, pp. 90--101.

\bibitem{DBLP:journals/corr/LayeghyPP16}
\BIBentryALTinterwordspacing
S.~Layeghy, F.~Pakzad, and M.~Portmann, ``{SCOR:} software-defined constrained
  optimal routing platform for {SDN},'' \emph{CoRR}, vol. abs/1607.03243, 2016.
  [Online]. Available: \url{http://arxiv.org/abs/1607.03243}
\BIBentrySTDinterwordspacing

\bibitem{moy1998ospf}
J.~T. Moy, \emph{{OSPF}: anatomy of an Internet routing protocol}.\hskip 1em
  plus 0.5em minus 0.4em\relax Addison-Wesley Professional, 1998.

\bibitem{gill2013survey}
P.~Gill, M.~Schapira, and S.~Goldberg, ``A survey of interdomain routing
  policies,'' \emph{ACM SIGCOMM Computer Communication Review}, vol.~44, no.~1,
  pp. 28--34, 2013.

\bibitem{Al-Fares:2008:SCD:1402958.1402967}
\BIBentryALTinterwordspacing
M.~Al-Fares, A.~Loukissas, and A.~Vahdat, ``A scalable, commodity data center
  network architecture,'' in \emph{Proceedings of the ACM SIGCOMM 2008
  Conference on Data Communication}, ser. SIGCOMM '08.\hskip 1em plus 0.5em
  minus 0.4em\relax New York, NY, USA: ACM, 2008, pp. 63--74. [Online].
  Available: \url{http://doi.acm.org/10.1145/1402958.1402967}
\BIBentrySTDinterwordspacing

\bibitem{optimization2014inc}
G.~Optimization, ``Inc.,“gurobi optimizer reference manual,” 2015,''
  \emph{URL: http://www. gurobi. com}, 2014.

\end{thebibliography}
}

\end{document}